\documentclass[a4paper,11pt]{article}
\usepackage[T1]{fontenc}
\usepackage[utf8]{inputenc}
\usepackage{lmodern}
\usepackage{graphicx}
\usepackage{epstopdf} 
\usepackage{color}
\title{A BRST view of the spontaneous symmetry breaking}
\author{R. L. P. G. Amaral$^{a}$\footnote{email: rubensamaral@id.uff.br} ,
	V. E. R. Lemes$^{b}$\footnote{email: verlemes@gmail.com},
	O. S. Ventura$^{c}$\footnote{email: ozemar.ventura@cefet-rj.br}, L.C.Q.Vilar $^{b}$\footnote{email: lcqvilar@gmail.com}  \\
	\small \em $^a$Instituto de F\'{\i}sica, Universidade Federal do Fluminense\\
	\small \em Av. Litor\^anea S/N, Boa Viagem, Niter\'oi-RJ CEP. 24210-340,
	Brazil\\
	\small \em $^b$Instituto de F\'\i sica, Universidade do Estado do Rio de
	Janeiro,\\
	\small \em Rua S\~{a}o Francisco Xavier 524, Maracan\~{a}, Rio de Janeiro - RJ,
	20550-013, Brazil\\
	\small \em $^c$Departamento de F\'\i sica, Centro Federal de Educa\c{c}\~ao Tecnol\'ogica do Rio de
	Janeiro\\
	\small\em Av.Maracan\~a 249, 20271-110, Rio de Janeiro - RJ, Brazil}

\begin{document}  

\maketitle
%\tableofcontents

\begin{abstract}
After the phase transition of a Grand Unified Theory (GUT), its single coupling splits into the couplings of the basic interactions. On the other side, the measurement of the Weinberg mixing angle in the eletroweak theory determines that the couplings associated to the SU(2) and U(1) symmetries were already distinct in its symmetric phase, before the electroweak breaking. This picture is consistent as we understand that such couplings evolve independently with the energy scale, meeting a common point together with the strong interaction coupling at the GUT scale. From a BRST symmetry point of view, this implies that each independent coupling should be associated to independent cocycles defined on a cohomological basis. The problem is that, before the GUT symmetry breaking,  the BRST operator associated to this symmetric phase predicts a single Yang-Mills coupling. This sets the question of what should be the BRST operator of the asymmetric phase, that would allow for the splitting of this coupling but at the same time remain compatible with the BRST operator of the symmetric phase. In this work we proceed to answer this question.
\end{abstract}

\section{Introduction}

The advent of a Grand Unified Theory (GUT) is the attempt to actually comprehend the standard model (SM) $SU(3)\times SU(2)\times U(1)$ physics inside a tied description of these basic interactions. The goal would be to reproduce all the SM richness from a theory with the smaller possible set of free parameters, bringing economy and a higher predictability to such a new physics. Some of its consequences are controversial, and, for example, the predicted decay of the proton present on the original model proposals \cite{GG74} and \cite{FM75,G75} is still under broad debate in the literature (see for instance the review \cite{NP07} or more recently \cite{FG17,FG18}). On the other side, these models establish in a simple way the connection of the electric charges of quarks and leptons. And mainly, the project of unification is supported by the experimental evidence that the strong, weak and electric couplings tend to converge on a common value at a high energy scale $M_G$ (in fact, this was the primary theoretical intention for the model design \cite{GQW74}). In this aspect, it is never excessive to remember that the electroweak theory in not a complete unification of the electric and weak interactions, as their couplings remain distinct even in the symmetric phase. This is also experimentally confirmed by the measurement of the Weinberg angle. Then, any GUT is ultimately a theory for the unification of all couplings, a theory with one single coupling. Our low energy physics would appear after a phase transition at the GUT scale $M_G$ by a Higgs mechanism, and since then the basic couplings evolved independently. This is understood as an effect of the decoupling of the heavy gauge modes which acquire masses proportional to $M_G$ at the new Higgs vacuum. This means that each one of these couplings follow an independent renormalization equation.

At this moment we make a brief interruption to state that the work that we will develop here is a rereading of the theoretical construction which is behind any GUT theory from the point of view of the BRST language. We think that this approach is most suitable, as the BRST charge should reflect the true invariance of the vacuum in each phase. Let us argue for this approach and how it can give us a new insight on this problem. 

Rephrasing our initial presentation, we would say that seeing from the standpoint of our low energy scale, it is the running of the couplings which allows the existence of a GUT theory. We expect them to meet at a definite high energy scale. They would join together in a unique coupling, which in its turn would then follow its own unique renormalization equation from this scale on. The SM group structure must then be accommodated within a simple Lie group. Now, from the perspective of the BRST analysis, this unique coupling is naturally predicted as a consequence of this character of the Lie group of the symmetric phase being simple.

 We can briefly summarize this line of reasoning and its impacts. In a first approximation, we can recover the concept of hidden symmetry, in contrast to a broken symmetry, originally exposed by Coleman \cite{Col85}. For this intent, it is sufficient just to shift the scalar field by its vacuum expectation value (vev), not only in the action, but also in the BRST operator of the symmetric phase. This will be the operator which will sign the transition from the symmetric to the broken phase. We will show that the immediate effect of this shift is to characterize the so acquired gauge mass in the usual cohomological BRST sense. Other consequences will be the trivialization of all cocycles containing any of the non derivated would be Goldstones, as they now form BRST doublets with the non derivated ghosts along the broken directions. This means that the physical Hilbert space do not depend on such fields (for this and other results in BRST cohomology we refer to \cite{Sor95}). The same will happen with certain combinations of the gauge fields associated to the broken generators  and the derivated would be Goldstones, as they now belong to BRST doublets with the derivatives of the same ghosts. As before, such combinations of fields will not belong to the physical space also. Comparing this result with a pure massless Yang-Mills theory, it is well known that in this case the BRST analysis leads to the physical independence on the non derivated gauge fields, and observables only depend on the fields curvatures \cite{Sor95}. Now, in the Higgs broken phase, this exclusion along the broken directions happens only in this specific combination. Consequently, we will show that, in the broken phase, the massive broken gauge fields will appear in non trivial BRST cocycles through the linear independent combinations with the would be Goldstones. In fact, this is responsible for the appearance of the Higgs gauge mass among the physical observables in the broken phase. And also this makes explicit what is the true physical variable in the broken gauge directions, which is nothing more than the well known statement that the would be Goldstones now represent the longitudinal degrees of freedom of the massive gauge fields in the broken Higgs phase. Then, from this BRST point of view, we are extending to the quantum level the classical observation that one can rotate the whole theory by a gauge transformation taking as gauge parameters the would be Goldstones (see for instance \cite{Wei96}). These scalar degrees of freedom disappear altogether from the theory as they are incorporated as the longitudinal massive gauge degrees.

We are now at this point where we have an understanding of the Higgs gauge mass, but we still lack the independence of the gauge couplings after symmetry breaking. We will show that this is so because the BRST operator that we extended by including the vev of the scalar field still predicts one single gauge coupling for the whole theory. There is only one possible non trivial element of the cohomology of this operator depending exclusively on the curvature, i.e. $Tr F^{2}$, and it is still based on the gauge field curvature of the symmetric phase. In this way, independent running gauge couplings after the phase transition are still not explained. 

Our proposal to approach this question is basically to reinterpret the quantization of the theory for the new Higgs vacuum. In this phase, as we have seen, the would be Goldstone fields (and their derivatives) do not play any physical role, as it happens with the ghosts (and derivatives) of the broken directions. So we proceed with the complete elimination of these fields from the theory, which can be understood as a choice of gauge, actually, the unitary gauge \cite{Wei96}. This will take us to a new BRST operator of the broken phase. This operator is nilpotent, and can be used to characterize a new cohomology for this phase. After showing that in this way we arrive at independent couplings for each independent symmetry of the new vacuum, we hope that this operator can guide us to the concept of observable after the symmetry breaking.

In the next section we will fix the notations and review some basic results of the BRST cohomology of massless Yang-Mills theory for completeness. In Section 3, we will show how we can characterize the emergence of the gauge field mass as a product of the change of the BRST operator driven by the vev of the scalar field. Also we will analyse the physical status of the would be Goldstone and of the broken ghost fields in view of this transition BRST operator. Here we will spend a few words on the 't Hooft gauge, responsible for simultaneously showing the renormalizability and unitarity of the theory, which is in fact the great advance of the Higgs mechanism for the introduction of masses for gauge fields. Then, in Section 4, we exhibit a nilpotent operator for the unitary quantization of the broken phase. We will study its cohomology at the level of the action, paying special attention to the non trivial elements depending exclusively on the gauge fields. This cohomology will present an independent running coupling for each remaining symmetry of the vacuum. The origin of each counterterm will be traced back to elements of the BRST cohomology of the symmetric phase. This will demand that  each independent running coefficient will have the necessary boundary condition of matching the associated value coming from the symmetric phase at the scale of the phase transition. This is the BRST derivation of the different gauge couplings matching  at the GUT scale $M_G$. At this point we will also show that the necessity of the 't Hooft gauge demands now the addition of a second independent symmetry, which will shape the full nilpotent BRST symmetry at the broken phase. This will fix the coefficients of the counterterm action. Finally, this development opens the possibility to incorporate a new physics, as new elements at the level of the action are allowed after the phase transition, elements that are associated to the symmetric phase cohomology in higher dimensions. The possible interpretation of such elements as condensate candidates will be commented in our Conclusions.

\section{Notations and BRST transformations}
We begin with the usual definitions for the curvature and covariant derivative of a Yang-Mills theory coupled to a scalar field
\begin{eqnarray}
F^{A}_{\mu \nu}&=&\partial_{\mu}A^{A}_{\nu}-\partial_{\nu}A^{A}_{\mu}+g f^{ABC}A^{B}_{\mu}A^{C}_{\nu} \, , \nonumber \\
(D_{\mu}\varphi)^{A}&=&\partial_{\mu}\varphi^{A}+gf^{ABC}A^{B}_{\mu}\varphi^{C}.
\end{eqnarray}
The gauge field $A^{A}_{\mu}$ and the scalar field  $\varphi^{A}$  are in the adjoint representation of the gauge group $G$, and then the internal index $A$ varies up to the dimension of $G$. We assume that this is a simple Lie group, which implies a single coupling constant $g$. The standard BRST transformations for these fields, ghost, curvature and covariant derivative are respectively given by
\begin{eqnarray}
sA^{A}_{\mu}&=& -(\partial_{\mu}c^{A}+gf^{ABC}A^{B}_{\mu}c^{C})= -(D_{\mu}c)^{A} \, , \nonumber \\
s\varphi^{A}&=& gf^{ABC}c^{B}\varphi^{C} \, , \nonumber \\
sc^{A}&=& \frac{g}{2}f^{ABC}c^{B}c^{C} \, , \nonumber \\
sF^{A}_{\mu \nu}&=&gf^{ABC}c^{B}F^{C}_{\mu \nu} \, , \nonumber \\
s(D_{\mu}\varphi)^{A}&=& gf^{ABC}c^{B}(D_{\mu}\varphi)^{C}.
\label{BRS}
\end{eqnarray}
This is the basic structure that we assume as that prevailing at the symmetric phase, prior to the phase transition. Before entering in the details of this process, we now briefly review how the cohomology of the nilpotent operator $s$ can be calculated \cite{Sor95}. These concepts will be sufficient to explore the broken phase. We begin by making a filtration in $s$,
\begin{eqnarray}
s&=& s_{0}+ s_{1} \, ,\nonumber \\
s^{2}&=& (s_{0})^{2}+ \{s_{0},s_{1}\}+ (s_{1})^{2} \, ,
\label{filter}
\end{eqnarray}
where $s_{0}$ acts linearly on the fields
\begin{eqnarray}
s_{0}A^{A}_{\mu}&=& -\partial_{\mu}c^{A}\, ,\nonumber \\
s_{0}c^{A}&=& 0\, ,\nonumber \\
s_{0}\varphi^{A}&=& 0\, .\nonumber \\
\label{s0}
\end{eqnarray}
The nilpotency of $s_{0}$ is easily confirmed from equation (\ref{s0}). It establishes that the gauge field $A^{A}_{\mu}$ and the derivative of the ghost $\partial_{\mu}c^{A}$ form what is called a $s_{0}$ doublet. A doublet $(u,v)$ of a nilpotent operator $\Delta$ happens when 
\begin{eqnarray}
\Delta u = v \quad , \quad \Delta v=0 \, ,
\label{DELTA}
\end{eqnarray}
and neither u nor v enter in any other field transformation of $\Delta$. Notice that, from  (\ref{s0}), the symmetrized derivative  $\partial_{(\mu}A^{A}_{\nu )}$ also belongs to a $s_{0}$  doublet, whereas $\partial_{[\mu}A^{A}_{\nu ]}$ is in a singlet . This filtration is useful thanks to two basic theorems \cite{Sor95}. The first one states that the cohomology of $s$ is isomorphic to a subspace of the cohomology of $s_{0}$. The second says that the cohomology of any nilpotent operator does not depend on doublets. Then, we can conclude that a basis for the cohomology of $s_{0}$ with any ghost number or dimension can be constructed from polynomials containing solely the nonderivated ghost, the scalar field and its derivatives, and/or the antisymmetrized gauge field derivative $\partial_{[\mu}A^{A}_{\nu ]}$. Expanding this basis with the quantum numbers of a four dimensional Lagrangian, and then using the first theorem to complete these elements to obtain the full cohomology of $s$, we arrive at the standard action of couterterms of Yang-Mills coupled to the scalar field. To each independent element on the cohomology of $s$, we associate a free coefficient. These coefficients are the physical parameters of the quantum action. From what we have just obtained, we can conclude that the only possible contribution to the counterterm Lagrangian made exclusively from gauge fields is $Tr F^{2}$. Then, the fact that polynomials constructed with the non-derivated gauge field $A^{A}_{\mu}$ are excluded from the physical space means that a gauge mass would be forbid as an observable even if an invariant cocycle could be built from them. As we know, one way to allow for gauge masses is to implement a symmetry breaking mechanism. In the following, we will show how this mechanism  circumvents this forbiddance from the BRST point of view.

\section{BRST at the phase transition}

The main purpose of this section is to show that we can define a BRST operator in a simple way in a phase transition. Spontaneous symmetry breaking does not mean that there is not a BRST symmetry regulating all the process, including the broken phase.

We suppose that this transition follows from the Higgs mechanism, when the scalar field acquires a vev $\mu$ along a direction $z$, 
\begin{equation}
 \varphi^{A}=\chi^{A}+\mu\delta^{Az}.
 \label{vev}
\end{equation}
We call $z$ the direction of the breaking. The shift in $\varphi$ implies $<\chi>=0$. We designate the directions in the internal space that commute with $z$ by lowercase letters from the middle of the alphabet, as i,j,k... These so called symmetric directions of the new vacuum may even form distinct blocks of $G$, with generators commuting among themselves. Of course, these blocks constitute subgroups of $G$, and are associated to the symmetries of the new vacuum. Moreover, in some equations, when necessary, we will use a subindex $(N)$ to refer to $N$ possible distinct subgroups. For example $T^{i(N)}$ will refer to the generator $i$ of the subgroup of label $N$. We also have the directions that do not commute with $z$, the so called broken directions, which we designate by lowercase letters from the beginning of the alphabet, as a,b,c... At this point, we specify the algebraic result of the broken theory by saying that we allow the structure constants of this phase to have only the following possible non-null contributions
\begin{eqnarray}
&&f^{zab},\hspace{3mm}f^{ijk},\hspace{3mm}f^{iab} \, .
\label{fs}
\end{eqnarray}
This may not be the most general possibility, but it is sufficient for our purposes (this follows the Cartan decomposition of a Lie algebra for a symmetric space when $f^{abc}$ vanishes \cite{Wei96}). For example, the breaking of $SU(5)$ into $SU(3)\times SU(2)\times SU(1)$, the original Georgi-Glashow model \cite{GG74}, follows this pattern. 

Then, at the moment of the breaking, the BRST transformation of the scalar field in (\ref{BRS}) must now take into account the vev $<\varphi^{z}>=\mu$, becoming the transition operator $s_{v}$
\begin{eqnarray}
s_{v}\chi^{A}&=& \mu g f^{ABz}c^{B} + g f^{ABC}c^{B}\chi^{C} \, ,
\end{eqnarray}
and from the list of available structure constants in (\ref{fs}) we get that only the scalar fields in the broken directions feel the presence of the vev,
\begin{eqnarray}
s_{v}\chi^{a}&=& \mu g f^{abz}c^{b} + g f^{abi}(c^{b}\chi^{i}- c^{i}\chi^{b}) 
+  g f^{abz}(c^{b}\chi^{z}-c^{z}\chi^{b})\, .
\label{chimu}
\end{eqnarray}
In the other directions, the scalar fields transformations remain unchanged
\begin{eqnarray}
s_{v}\chi^{i}&=& g f^{ijk}c^{j}\chi^{k} + g f^{iab}c^{a}\chi^{b} \, ,\nonumber \\
s_{v}\chi^{z}&=& g f^{zab}c^{a}\chi^{^{b}}.
\label{chiz}
\end{eqnarray}
Here it is important to highlight that the operator $s_{v}$ is in fact a symmetry of the whole action after the symmetry breaking, being understood that for the other fields $s_{v}$ acts as $s$ in (\ref{BRS})
\begin{eqnarray}
s_{v}A^{A}_{\mu}&=& -(\partial_{\mu}c^{A}+gf^{ABC}A^{B}_{\mu}c^{C})= -(D_{\mu}c)^{A} \, , \nonumber \\
s_{v}c^{A}&=& \frac{g}{2}f^{ABC}c^{B}c^{C} \, .
\label{sv}
\end{eqnarray}
This can be seen as an explicit realization of the Coleman's "hidden symmetry"  description of a broken symmetry \cite{Col85}. But the change for this hidden symmetry $s_{v}$ obviously has  physical consequences. 

The novelty comes from the shift now present in the transformation (\ref{chimu}). It has the power to completely change the cohomology. In order to understand this, we can filter $s_{v}$  as in (\ref{filter}), and obtain $s_{0v}$ ,
\begin{eqnarray}
s_{0v}A^{A}_{\mu}&=& -\partial_{\mu}c^{A} \, ,\nonumber \\
s_{0v}c^{A}&=& 0 \, ,\nonumber \\
s_{0v}\chi^{i}&=& 0 \, ,\nonumber \\
s_{0v}\chi^{z}&=& 0 \, ,\nonumber \\
s_{0v}\chi^{a}&=& \mu g f^{abz}c^{b} \, .
\label{s0v}
\end{eqnarray}
From (\ref{s0v}) we immediately see that the scalars $\chi^{a}$ are now in doublets with the ghosts $c^{a}$ arranged in linearly independent combinations with coefficients given by the $f^{abz}$ (here we are implicitly supposing that the matrix $M^{ab}=f^{abz}$ does not have zero modes). Besides this, we can see that the combination $\mu g f^{abz}A^{b}_{\mu}-\partial_{\mu}\chi^{a}$ belongs to a doublet with $\mu g f^{abz}\partial_{\mu}c^{b}$. At the same time, the linearly independent combination $\mu g f^{abz}A^{b}_{\mu}+\partial_{\mu}\chi^{a}$ is a singlet,
\begin{equation}
s_{0v}( \mu g f^{abz}A^{b}_{\mu}+\partial_{\mu}\chi^{a})=0 \, .
\label{singlet}
\end{equation}
Then, using the cohomological doublets theorem stated in the last section, the basis for the cohomology of $s_{0v}$, limited in four dimensions, is expanded by polynomials constructed from this combination presented in (\ref{singlet}), the curvatures $\partial_{[\mu}A^{A}_{\nu ]}$, and the scalars, ghosts and their derivatives still belonging to $s_{0v}$ singlets, in particular all the fields $\chi^{i}$ and $\chi^{z}$ (and derivatives), and the nonderivated ghosts $c^{i}$ and $c^{z}$.

The presence of the element in (\ref{singlet}) in the cohomology opens the possibility of the existence of monomials of dimension 2 in the action with bilinears in the $A^{a}$. This just means that these vectorial bosons which enter the singlets in (\ref{singlet}) acquire masses after the symmetry breaking. At the same time, we interpret the scalars $\chi^{a}$ whose derivatives appear in (\ref{singlet}) as the would be Goldstone bosons. As they themselves belong to doublets, we understand that none physical observable will depend on them on the broken phase.

It is not difficult to show that (\ref{singlet}) can be completed to the cohomology of the full operator $s_{v}$ 
\begin{eqnarray}
V^{A}_{\mu}&=& \mu g f^{Abz}A^{b}_{\mu}+(D_{\mu}\chi)^{A} \, , 
\label{VA}
\end{eqnarray}
transforming covariantly as
\begin{eqnarray}
s_{v}V^{A}_{\mu}&=& g f^{ABC}c^{B}V^{C}_{\mu} \, .
\end{eqnarray}
This kind of relation was called a "Russian-like formula" in \cite{Car95}, after its functional similarity with the Russian formula (see \cite{Sor95} and references therein). Basically, it represents a mapping of the cohomology of a filtered operator to the cohomology of the full nilpotent operator. In our present case, this relation will enable us to concentrate our efforts of the cohomology analysis in the next sections to the sector built exclusively with gauge fields.

Then, remembering the allowed $f^{ABC}$ given in (\ref{fs}),
\begin{eqnarray}
V^{A}_{\mu}V^{A\mu} &=& \mu^{2}g^{2} f^{abz}f^{acz}A^{b}_{\mu} A^{c\mu} +2g\mu f^{abz}A^{b}_{\mu}(D\chi)^{a\mu} + (D\chi)^{A}_{\mu}(D\chi)^{A\mu}
\label{V2}
\end{eqnarray}
is the invariant nontrivial cocycle of the cohomology of the transition BRST operator  $s_{v}$. The first term in (\ref{V2}) is the massive contribution just mentioned, and the second contains the well known nonphysical coupling of the massive gauge fields and the would be Goldstones.

Here it is important to dwell on the analysis of this point. As we have shown, the fact that each would be Goldstone becomes a nonphysical representation follows essentially from their transition BRST transformations (\ref{chimu}). Actually, this resembles the same problem of the Yang-Mills action in the symmetric phase, with the explicit presence of a longitudinal mode of the gauge field which does not represent a physical degree of freedom. This is solved by the existence of the BRST symmetry itself, which enables a gauge fixing procedure to eliminate this mode. From another point of view, the invariance under a gauge symmetry implies that not all mathematically possible gauge connections are physically distinguishable, and again the question of a gauge fixing is central. Then, on the one hand a gauge fixing is demanded by the gauge symmetry, on the other it is allowed by the BRST symmetry. The Faddev-Popov mechanism follows from these principles. 

Alternatively, the mechanism for the case of a spontaneous symmetry breaking was designed by 't Hooft \cite{Hoo71,Fuj72}. Its importance relies not only on the elimination of the nonphysical couplings shown in (\ref{V2}), but also for a limit where renormalizability becomes explicit, depending only on the choice of a gauge parameter. This ensures unitarity and renormalizabilty at the same time. This is essential for the theoretical framework of the Higgs mechanism. And it is the existence of the BRST symmetry that states that the choice of any specific value of the gauge parameter is physically irrelevant (physical observables do not dependend on this choice \cite{Sor95}).  We can compare with the situation of a Yang-Mills theory with a mass term. In this case, as the gauge symmetry is explicitly broken, we loose the BRST symmetry altogether. This implies that it is not possible to implement a gauge fixing, and this theory is inevitably power-counting nonrenormalizable (there is an elucidative argument in \cite{Itz80} on this point).

Coming back to the nonphysical coupling in (\ref{V2}), we can introduce the term proposed by 't Hooft
\begin{equation}
G^{A}=\partial_{\mu}A^{A\mu} + \alpha \mu g f^{Azb}\chi^{b} \, ,
\label{ga}
\end{equation}
together with a $s_{v}$ doublet composed of an anti-ghost $q^{A}$ and a Lagrange multiplier $b^{A}$ transforming as
\begin{eqnarray}
s_{v}q^{A}=b^{A}, \hspace{5mm} s_{v}b^{A}=0. 
\label{svqa}
\end{eqnarray}
The 't Hooft gauge fixing can now be written as a transition BRST trivial cocycle
\begin{eqnarray}
S_{gf}&=& s_{v}\int d^{4}x (2q^{A}G^{A}-\alpha q^{A}b^{A}) \, ,
\label{sgf}
\end{eqnarray}
where $\alpha$ is the gauge parameter just mentioned.
When we integrate out the multiplier $b^{A}$, the spurious nonphysical element in (\ref{V2}) is eliminated. And it is the triviality of (\ref{sgf}), possible by the existence of the transition BRST  $s_{v}$ as a symmetry of the action, that is behind the independence of the physics on $\alpha$. An adequate choice of the gauge parameter then makes the massive propagator compatible with power counting renormalizability. Once physics is independent on this choice, the theory is renormalizable in any gauge.

So in (\ref{V2}) we understood how a phase transition can change the cohomology. New elements can appear in the broken phase, which were not even BRST symmetric in the previous phase. However, this still cannot be the complete description. As we said, all this development pursuits the construction of a GUT, with the evidence that the single coupling of the symmetric phase must split into the independent couplings of the basic interactions at our energy scale. The problem is that following the standard calculation, the cohomology of $s_{v}$ in the sector made exclusively of gauge fields is identical to the cohomology of $s$, i.e., there is still only one cocycle, the same $F^{A}_{\mu\nu}F^{A\mu\nu}$, and then only one possible coupling. The question is how we can modify or reinterpret the cohomology after the symmetry breaking scale in a way compatible with this transition between both regimes. This is the issue for the next section

\section{Effective BRST Symmetries of The New Vacuum}

We begin this section remembering that from (\ref{s0v}) we concluded that the ghosts $c^{a}$  and the scalars $\chi^{a}$ belong to the trivial sector of the theory after the symmetry breaking, as it happens to all ghost derivatives. Actually, the would be Goldstones only appear in the combination given in (\ref{VA}), as part of the massive gauge field, the longitudinal component. This is in agreement with the classical argument that these Goldstones can be rotated away, being completely eliminated from the theory. This is equivalent to a gauge choice, the unitary gauge \cite{Wei73}. Then, from this point on, we will restrict our analysis to the sector built exclusively by the original gauge field $A^{A}_{\mu}$  (soon, when we limit our cohomology study to the level of the 4D couterterm action, we will understand that wherever $A^{a}_{\mu}$ appears non-derivated it must be completed in the sense of the Russian-like formula (\ref{VA})). The gauge freedom associated to the symmetries of the new vacuum, due to the non null $f^{ijk}$, can be shown to be preserved in this process, and need to be gauge fixed \cite{Wei73}. And although the gauge components $A^{a}_{\mu}$ are now massive, a gauge fixing is still needed along these directions, as the discussion on 't Hooft gauge of the last section specified. This means that we cannot eliminate the ghosts $c^{a}$ altogether, but we can isolate them from the BRST machinery in the first approach. In addition, we still have the necessary link with the theory at the breaking point determined by the transition BRST  $s_{v}$. Then, joining all these considerations, the BRST quantization for the broken phase can start from the filtration of the fields  $c^{a}$ from eqs. (\ref{sv}) (see \cite{Sor95} for details on this mathematical step). Noticing the allowed structures constants (\ref{fs}) for the symmetric phase, the new BRST operator $s_{q}$ becomes
\begin{eqnarray}
s_{q}A^{i(N)}_{\mu}&=& -\partial_{\mu}c^{i(N)}- g_{N}f^{ijk(N)}A^{j(N)}_{\mu}c^{k(N)} \, , \nonumber \\
s_{q}A^{z}_{\mu}&=& -\partial_{\mu}c^{z} \, , \nonumber \\
s_{q}A^{a}_{\mu}&=& - \sum_{N} g_{N}f^{abi(N)}A^{b}_{\mu}c^{i(N)} - g^{\prime} f^{abz}A^{b}_{\mu}c^{z} \, , \nonumber \\
s_{q}\chi^{i(N)}&=&g_{N}f^{ijk(N)}c^{j(N)}\chi^{k(N)} \, , \nonumber \\
s_{q}\chi^{z}&=&0 \, , \nonumber \\
s_{q}\chi^{a}&=& - \sum_{N} g_{N}f^{abi(N)}\chi^{b}c^{i(N)} - g^{\prime} f^{abz}\chi^{b}c^{z} \, , \nonumber \\
s_{q}c^{i(N)}&=&\frac{1}{2} g_{N}f^{ijk(N)}c^{j(N)}c^{k(N)} \, , \nonumber \\
s_{q}c^{z}&=&0 \, , \nonumber \\
s_{q}c^{a}&=& \sum_{N} g_{N}f^{abi(N)}c^{b} c^{i(N)} + g^{\prime} f^{abz}c^{b} c^{z}.
\label{sq}
\end{eqnarray}
The main point now that one can check is that the set of transformations (\ref{sq}) define a nilpotent $s_{q}$ operator in all the fields of the theory indeed. This is fundamental in order to view this set as a BRST system.
In eq. (\ref{sq}) we finally needed to specify the $N$ possible different subgroups of the original gauge group that remain as symmetries of the new vacuum. Then $A^{i(N)}$ means the $i$ component of the gauge field associated to the subgroup $N$. Independence of each subgroup means that the structure constant $f^{ijk(N)}$ is only non vanishing if all its indices belong to the same subgroup $N$, so that if $T^{i(N)}$ is a generator associated to one of these subgroups, then 
\begin{equation}
[T^{i(N)},T^{j(M)}]=i\delta^{(N)(M)}f^{ijk(N)}T^{k(N)} \, .
\end{equation}
The label $N$ in $f^{abi(N)}$ entering the $s_{q}$ transformation of $A^{a}_{\mu}$ in (\ref{sq}) means that the index $i$ belongs to the subgroup $N$. Summation over this label $N$ will always be explicitly indicated, otherwise it is not intended. This happens, for example, in the transformation of $A^{i(N)}$ in (\ref{sq}), where no summation on $N$ occurs. 

Finally, we call attention to the distinct couplings $g_{N}$ introduced for each subgroup and the coupling  $g^{\prime}$ associated to the abelian subgroup along the symmetry breaking direction $z$. The imposition that the system (\ref{sq}) must match that formed by (\ref{chimu}), (\ref{chiz}) and (\ref{sv}) imply that all these couplings 
$g_{N}$  and  $g^{\prime}$ must satisfy the boundary condition of matching $g$, the coupling in the symmetric phase, at the GUT symmetry breaking scale
\begin{equation}
g_{N}(M_{G})= g^{\prime}(M_{G})= g(M_{G}) \, .
\label{gs}
\end{equation}
Obviously, the equivalence among the couplings cannot be preserved at our energy scale. This indicates that they must evolve independently after the GUT phase transition, which means that each one must be renormalized following its own renormalization equation \cite{Wei96}. In the BRST quantization procedure, this implies that each coupling must be associated to an independent nontrivial cocycle in the cohomology of the BRST operator at this phase, which is again the problem that we were facing at the end of the last section. But now, the BRST $s_{q}$ for the broken phase gives us the answer. In fact, it is immediate to confirm that the cohomology with the quantum numbers of the four dimensional action calculated from the BRST transformations (\ref{sq}) has  independent contributions of the form
\begin{eqnarray}
F^{i(N)}_{\mu\nu}F^{i(N)\mu\nu} \, ,
\label{FiN}
\end{eqnarray}
where
\begin{eqnarray}
F^{i(N)}_{\mu\nu}&=& \partial_{\mu}A^{i(N)}_{\nu}-\partial_{\nu}A^{i(N)}_{\mu}+ g_{N}f^{ijk(N)}A^{j(N)}_{\mu}A^{k(N)}_{\nu} \, .
\end{eqnarray}
We will also have the nontrivial cocycle
\begin{eqnarray}
F^{z}_{\mu\nu}F^{z\mu\nu} \, ,
\label{Fz}
\end{eqnarray}
with
\begin{eqnarray}
F^{z}_{\mu\nu}&=& \partial_{\mu}A^{z}_{\nu}-\partial_{\nu}A^{z}_{\mu}\nonumber \\
\end{eqnarray}
associated to an independent abelian coupling $g^{\prime}$. This set of contributions (\ref{FiN}) and (\ref{Fz}) indicates independent beta functions to each coupling of the independent subgroups that remain as symmetries of the vacuum after the phase transition.

Once we have obtained the cohomological independence of the physical couplings, we must stress that up to this point this has just served us as a guideline to develop the correct BRST system adequate to the experience at our energy scale. From this point on, we will explore some possible new physics that may be achieved by the study of the system (\ref{sq}). In this work we will concentrate on the rest of the cohomology at the level of the action built exclusively from the vector fields. All terms in this sector which are independent of the massive $A^{a}_{\mu}$ are described by (\ref{FiN}) and (\ref{Fz}). In the broken directions we may see some changes. If we apply the same filtration done in eq. (\ref{s0}) to the operator $s_{q}$, it indicates that  $A^{a}_{\mu}$ becomes a singlet of this filtered operator. In order to expand its cohomology upon this sector, we must include any object built with $A^{a}_{\mu}$ and its derivatives. The most general polynomial in this sector with the quantum numbers of the action is

\begin{eqnarray}
\Sigma &=& \int d^{4}x (a_{1}\partial_{\mu}A^{a}_{\nu}\partial^{\mu}A^{a\nu} + 
a_{2}\partial_{\mu}A^{a}_{\nu}\partial^{\nu}A^{a\mu} +
a_{3}^{ab}A^{a}_{\mu}A^{b\mu} + a_{4}(A^{a}_{\mu}A^{a\mu})^{2} + a_{5}A^{a}_{\mu}A^{a}_{\nu}A^{b\mu}A^{b\nu}\nonumber \\
&+&\sum_{N} a_{6}^{(N)}f^{abi(N)}A^{a\mu}A^{b\nu}f^{cdi(N)}A^{c}_{\mu}A^{d}_{\nu} 
+ a_{7}f^{zab}A^{a\mu}A^{b\nu}f^{zcd}A^{c}_{\mu}A^{d}_{\nu}
+ \sum_{N}a_{8}^{(N)}F^{i(N)}_{\mu\nu}f^{abi(N)}A^{a\mu}A^{b\nu}\nonumber \\
&+& a_{9}F^{z}_{\mu\nu}f^{zab}A^{a\mu}A^{b\nu}
 + D^{abcd}A^{a\mu}A^{b\mu}A^{c\nu}A^{d\nu} )\, ,
\label{sigma}
\end{eqnarray}
where by $D^{abcd}$ we mean all possible independent symmetric elements with four indices in the gauge algebra. For example, depending on the GUT group, even the quartic symmetric symbol $d^{abcd}$ obtained from the fully symmetric trace of four generators is allowed (see \cite{Que18} for details on this symbol). They will also depend on the representation used, and their detailed analysis is beyond the scope of this work. Anyway, we will give an example of what we mean by independence of these terms in the end of this section.

We will soon discuss each element written in (\ref{sigma}) and search for their completion in the cohomology of the full $s_{q}$. First we notice the presence of the massive contribution with coefficient $a_{3}$. It is a nontrivial contribution of the cohomology of  $s_{q}$, but as previously emphasized, it should be understood as part of the invariant cocycle (\ref{V2}). On the other hand, the elements with coefficients $a_{4}$ , $a_{5}$ and $D^{abcd}$, also immediately invariant under $s_{q}$, cannot be understood as coming from elements already present at the original action of the symmetric phase. When we come to the definition of the counterterm action this aspect will just be responsible for their exclusion, but at this point their presence means the possibility of a new physics as long as the  $s_{q}$ operator can be associated to an effective symmetry of the vacuum at a low energy scale. We will have more to say about this later in this section and in the Conclusions.

So, let us complete the other elements in (\ref{sigma}) to obtain the cohomology of  $s_{q}$. This problem can be seen from the point of view of a consistent deformation \cite{Hen93}. We will show in details an example of this process. We can begin with the first element by calculating its variation under $s_{q}$,
\begin{eqnarray}
s_{q} \int d^{4}x (\partial_{\mu}A^{a}_{\nu}\partial^{\mu}A^{a\nu}) = - \int d^{4}x (2\sum_{N}g_{N}\partial_{\mu}(f^{abi(N)}A^{b}_{\nu}c^{i(N)})\partial^{\mu}A^{a\nu} + 2g^{\prime}\partial_{\mu}(f^{abz}A^{b}_{\nu})c^{z}\partial^{\mu}A^{a\nu}) \, .
\label{sqdada}
\end{eqnarray}
The element of first order in the couplings whose $s_{q}$ transformation can compensate 
these in (\ref{sqdada}) is
\begin{equation}
-\int d^{4}x (\sum_{N}g_{N}f^{abi(N)}A^{i(N)}_{\mu}A^{b}_{\nu}\partial^{\mu}A^{a\nu} + g^{\prime}f^{abz}A^{z}_{\mu}A^{b}_{\nu}\partial^{\mu}A^{a\nu}) \, .
\label{dada2}
\end{equation}
In its turn, the $s_{q}$ variation of this element gives a second order contribution in the couplings which can be compensated by the introduction of the following second order element
\begin{eqnarray}
\int d^{4}x (\sum_{N}\sum_{M}g_{N}g_{M}f^{abi(N)}f^{acj(M)}A^{i(N)}_{\mu}A^{j(M)}_{\mu}A^{b}_{\nu}A^{c}_{\nu} 
+2\sum_{N}g_{N}g^{\prime}f^{abi(N)}f^{acz}A^{i(N)}_{\mu}A^{z}_{\mu}A^{b}_{\nu}A^{c}_{\nu} \nonumber \\
+(g^{\prime})^{2}f^{azb}f^{azc}A^{z}_{\mu}A^{z}_{\mu}A^{b}_{\nu}A^{c}_{\nu}) \, .
\label{dada3}
\end{eqnarray}
If we join together the first element in (\ref{sigma}) with those in (\ref{dada2}) and (\ref{dada3}) we can see the formation of a covariant derivative
\begin{eqnarray}
(\nabla_{\mu}A_{\nu})^{a}&=&\partial_{\mu}A^{a}_{\nu}-\sum_{N} g_{N}f^{abi(N)}A^{i(N)}_{\mu}A^{b}_{\nu}-g^{\prime}f^{abz}A^{z}_{\mu}A^{b}_{\nu} \, 
\label{dercov}
\end{eqnarray}
where covariance here means that it transforms as the $A^{a}_{\mu}$ field under $s_{q}$ as shown in (\ref{sq})
\begin{eqnarray}
s_{q}(\nabla_{\mu}A_{\nu})^{a} = 
- \sum_{N} g_{N}f^{abi(N)} (\nabla_{\mu}A_{\nu})^{b}c^{i(N)} - 
g^{\prime} f^{abz}(\nabla_{\mu}A_{\nu})^{b}c^{z} \, .
\label{sqnabla}
\end{eqnarray}
Then, the nontrivial cocycle in the cohomology of $s_{q}$ constructed from the element $a_{1}$ in (\ref{sigma}) assumes the simple form
\begin{equation}
 a_{1}\int d^{4}x (\nabla_{\mu}A_{\nu})^{a}(\nabla^{\mu}A^{\nu})^{a} \, .
 \label{a1}
\end{equation}
Equations (\ref{dercov}) and (\ref{a1}) indicate that $A^{a}_{\mu}$  is in fact a vectorial matter field in the broken phase. Notice, however, that it is not in any representation of the invariant groups of the new vacuum. As we can see in (\ref{sq}), its transformation uses the structure constants of the symmetric phase in the broken directions, which appear again in the definition of the covariant derivative (\ref{dercov}). This seems to be related to a necessity of recovering the full theory in the symmetric phase. Let us show how this may happen. 

Before the GUT symmetry breaking, we had the contribution in the Yang-Mills action 
\begin{eqnarray}
\int d^{4}x ( F^{a}_{\mu\nu}F^{a\mu\nu}) = \int d^{4}x (\partial_{\mu}A^{a}_{\nu}-\partial_{\nu}A^{a}_{\mu}+g f^{aBC}A^{B}_{\mu}A^{C}_{\nu})(\partial^{\mu}A^{a\nu}-\partial^{\nu}A^{a\mu}+g f^{aDE}A^{D\mu}A^{E\nu}) \, .
\label{fafa}
\end{eqnarray}
Notice that in the right hand side of (\ref{fafa}) we are only adding on the broken directions of the Yang-Mills curvature, this is not the full Yang-Mills action yet. If we substitute the possible structure constants (\ref{fs}) after the symmetry breaking
we get
\begin{eqnarray}
\int d^{4}x ( F^{a}_{\mu\nu}F^{a\mu\nu}) = \int d^{4}x (\partial_{\mu}A^{a}_{\nu}-\partial_{\nu}A^{a}_{\mu}-\sum_{N} g_{N}f^{abi(N)}A^{i(N)}_{\mu}A^{b}_{\nu}+\sum_{N} g_{N}f^{abi(N)}A^{b}_{\mu}A^{i(N)}_{\nu} \nonumber \\
-g^{\prime}f^{abz}A^{z}_{\mu}A^{b}_{\nu}+g^{\prime}f^{abz}A^{b}_{\mu}A^{z}_{\nu})^{2} \, .
\end{eqnarray}
Finally, after (\ref{dercov}) it is immediate to see that
\begin{eqnarray}
\int d^{4}x ( F^{a}_{\mu\nu}F^{a\mu\nu}) = \int d^{4}x (\nabla_{\mu}A_{\nu}-\nabla_{\nu}A_{\mu} )^{a}(\nabla^{\mu}A^{\nu}-\nabla^{\nu}A^{\mu})^{a} \, .
\end{eqnarray}
We identify the origin of the cocycle (\ref{a1}) as part of the Yang-Mills action before the GUT symmetry breaking. The observation is that at this point of the analysis  of the broken phase, we now have independent coefficients for each contraction, as can be understood from (\ref{sigma}) and (\ref{a1})
\begin{equation}
 \int d^{4}x (a_{1}(\nabla_{\mu}A_{\nu})^{a}(\nabla_{\mu}A_{\nu})^{a} + a_{2}(\nabla_{\mu}A_{\nu})^{a}(\nabla^{\nu}A^{\mu})^{a}) \, .
 \label{a1a2}
\end{equation}

Let us continue now with the other elements of (\ref{sigma}). In order to study the elements $a_{7}$ and  $a_{9}$, we define $F^{A=z}_{\mu\nu}$ as the component $z$ of $F^{A}_{\mu\nu}$
\begin{eqnarray}
F^{A=z}_{\mu\nu}&=& \partial_{\mu}A^{z}_{\nu}-\partial_{\nu}A^{z}_{\mu}+gf^{zab}A^{a}_{\mu}A^{b}_{\nu} \, .
\end{eqnarray}
Since
\begin{eqnarray}
sF^{A=z}_{\mu\nu} &=&- gf^{zab}F^{a}_{\mu\nu}c^{b} \, ,
\end{eqnarray}
it implies
\begin{eqnarray}
s_{q}F^{A=z}_{\mu\nu}&=&0 \, ,
\end{eqnarray}
and then it is obvious that
\begin{equation}
s_{q}(\partial_{\mu}A^{z}_{\nu}-\partial_{\nu}A^{z}_{\mu})=0
\Longrightarrow s_{q}(f^{zab}A^{a}_{\mu}A^{b}_{\nu})=0 \, .
\label{sqfz}
\end{equation}
From the last inference in (\ref{sqfz}), we see that the terms  $a_{7}$ and  $a_{9}$  of (\ref{sigma}) are already invariant under $s_{q}$.

For the last elements in (\ref{sigma}), and following the same reasoning as before, we see that
\begin{eqnarray}
s_{q}F^{A=i(N)}_{\mu\nu} &=& g_{N}f^{ijk(N)}c^{j(N)}F^{A=k(N)}_{\mu\nu} \, ,
\label{sqfain}
\end{eqnarray}
with
\begin{eqnarray}
F^{A=i(N)}_{\mu\nu}&=& \partial_{\mu}A^{i(N)}_{\nu}-\partial_{\nu}A^{i(N)}_{\mu}+g_{N}f^{ijk(N)}A^{j(N)}_{\mu}A^{k(N)}_{\nu}+g_{N}f^{abi(N)}A^{a}_{\mu}A^{b}_{\nu}\nonumber \\
&=&F^{i(N)}_{\mu\nu}+g_{N}f^{abi(N)}A^{a}_{\mu}A^{b}_{\nu} \, ,
\label{fain}
\end{eqnarray}
where we used the definition (\ref{FiN}) for $F^{i(N)}_{\mu\nu}$. From (\ref{sq}), we know that
\begin{eqnarray}
s_{q}F^{i(N)}_{\mu\nu} = g_{N}f^{ijk(N)}c^{j(N)}F^{k(N)}_{\mu\nu} \, .
\label{sqfin}
\end{eqnarray}
Then, substituting (\ref{fain}) and (\ref{sqfin}) in (\ref{sqfain}), we arrive at the covariant transformation that we were searching
\begin{equation}
s_{q}(f^{abi(N)}A^{a}_{\mu}A^{b}_{\nu})=g_{N}f^{ijk(N)}c^{j(N)}(f^{abk(N)}A^{a}_{\mu}A^{b}_{\nu}) \, .
\end{equation}
From this, one easily obtain the invariance of the last elements $a_{6}$ and $a_{8}$ under $s_{q}$. This finishes the proof that each coefficient in (\ref{sigma}) is associated to an invariant nontrivial cocycle of $s_{q}$.

More than that, we see that if we join the terms $a_{6}$, $a_{7}$, $a_{8}$, and  $a_{9}$ of (\ref{sigma}),  together with the elements (\ref{FiN}), (\ref{Fz}), and (\ref{a1a2}) that we have already proved to be in the cohomology of $s_{q}$, we recover the full structure of the original Yang-Mills term $F^{A}_{\mu\nu}F^{A\mu\nu}$ of the symmetric phase. What this is in fact proving to us is that the effect of the symmetry breaking is to split the original cocycles of the symmetric phase into independent invariant nontrivial pieces after the breaking. 

Gathering togheter the nontrivial contributions (\ref{FiN}), (\ref{Fz}), (\ref{a1a2}), the already invariant elements $a_{3}$ to $a_{9}$ and $D_{abcd}$ from (\ref{sigma}), we display the cohomology of $s_{q}$ at the broken phase in the sector of the gauge fields
\begin{eqnarray}
 \Sigma_{c} & = &\int  d^{4}x(a^{\prime}F^{z}_{\mu\nu}F^{z\mu\nu} + \sum_{N} a_{N}F^{i(N)}_{\mu\nu}F^{i(N)\mu\nu}+a_{1}(\nabla_{\mu}A_{\nu})^{a}(\nabla_{\mu}A_{\nu})^{a} + a_{2}(\nabla_{\mu}A_{\nu})^{a}(\nabla^{\nu}A^{\mu})^{a}+a_{3}^{ab}A^{a}_{\mu}A^{b\mu} \nonumber \\
&+& a_{4}(A^{a}_{\mu}A^{a\mu})^{2} + a_{5}A^{a}_{\mu}A^{a}_{\nu}A^{b\mu}A^{b\nu}+ \sum_{N}a_{6}^{(N)}f^{abi(N)}A^{a\mu}A^{b\nu}f^{cdi(N)}A^{c}_{\mu}A^{d}_{\nu} 
+ a_{7}f^{zab}A^{a\mu}A^{b\nu}f^{zcd}A^{c}_{\mu}A^{d}_{\nu}
\nonumber \\
&+& \sum_{N}a_{8}^{(N)}F^{i(N)}_{\mu\nu}f^{abi(N)}A^{a\mu}A^{b\nu}+ a_{9}F^{z}_{\mu\nu}f^{zab}A^{a\mu}A^{b\nu}+ D^{abcd}A^{a\mu}A^{b\mu}A^{c\nu}A^{d\nu} ) \, .
\label{sigmac}
\end{eqnarray}

Then we see from (\ref{sigmac}) that $s_{q}$ leaves several free coefficients. Actually this is not a desired feature.
Following the BRST quantization procedure, for each of these  independent coefficients we would associate a physical observable of the broken phase. For example, independent elements as $a_{1}$ and $a_{2}$ do not seem reasonable. The clue to the missing piece is to recover a point mentioned in the beginning of this section: a gauge fixing is still needed along the broken directions. This means that we need a propagation for the ghost $c^{a}$, as can be obtained from the expression (\ref{sgf}) for the 't Hooft gauge fixing. The problem that we face now is that once the operator becomes $s_{q}$ instead of $s_{v}$ in (\ref{sgf}), we loose this propagator. This signs to us that $s_{q}$ cannot be the whole BRST operator of the vacuum after the symmetry breaking. Something must be missing and $s_{q}$ should be completed with extra terms. Here we call this missing piece as the operator $\delta$, and comparing the $s_{q}$ operator of (\ref{sq}) with the $s_{v}$ of  (\ref{sv}), we guess that $\delta$ in general can be written as

\begin{eqnarray}
\delta A^{i(N)}_{\mu}&=& - X_{N}f^{iab(N)}A^{a}_{\mu}c^{b} \, , \nonumber \\
\delta A^{z}_{\mu}&=& - X_{z}f^{zab}A^{a}_{\mu}c^{b} \, , \nonumber \\
\delta A^{a}_{\mu}&=& - \sum_{N} Y_{N}f^{aib(N)}A^{i(N)}_{\mu}c^{b} -  Y_{z} f^{azb}A^{z}_{\mu}c^{b} -\partial_{\mu}c^{a} \, , \nonumber \\
\delta \chi^{i(N)}&=&-L_{N}f^{iab(N)}\chi^{a}c^{b} \, , \nonumber \\
\delta \chi^{z}&=&-L_{z}f^{zab}\chi^{a}c^{b} \, , \nonumber \\
\delta \chi^{a}&=& - \sum_{N} K_{N}f^{aib(N)}\chi^{i(N)}c^{b} -  K_{z} f^{azb}\chi^{z}c^{b} - M_{z} \mu f^{abz}c^{b} \, , \nonumber \\
\delta c^{i(N)}&=&\frac{1}{2} B_{N}f^{iab(N)}c^{a}c^{b} \, , \nonumber \\
\delta c^{z}&=&\frac{1}{2} B_{z}f^{zab}c^{a}c^{b} \, , \nonumber \\
\delta c^{a}&=& 0.
\label{delta}
\end{eqnarray}

The coefficients $X_{N}$, $X_{z}$, $Y_{N}$, $Y_{z}$, $L_{N}$, $L_{z}$, $K_{N}$, $K_{z}$, $M_{z}$, $B_{N}$, $B_{z}$ are fixed by demanding that the full operator $s_{q}+\delta$ be nilpotent on all the set of fields of the theory. This constraint has a solution for these coefficients as

\begin{eqnarray}
	X_{N}= B_{N}=\frac{1}{Y_{N}}=\frac{1}{g_{N}} \, , \nonumber \\
	X_{z}= B_{z}=\frac{1}{Y_{z}}=\frac{1}{g^{\prime}}
	, \nonumber \\
	L_{N}.K_{N}=L_{z}.K_{z}=-M_{z}= 1 . 
	\label{coef}
\end{eqnarray}

This allows us to interpret $s_{q}+\delta$ as the adequate BRST operator for the theory after the phase transition. Although an interesting point to be observed is that the dependence on the inverse of the non-abelian couplings $g_{N}$ suggests that $\delta$ would loose its relevance on lower energy scales, where the theory would be dominated by the $s_{q}$ part of the BRST symmetry.

Once we obtain $\delta$, we must impose it as a symmetry of the action (\ref{sigmac}). Finally, the free coefficients of (\ref{sigmac}) are fixed as

\begin{eqnarray}
	a_{N}= \frac{g_{N}^2}{4} \, , \nonumber \\
	a^{\prime}=\frac{g^{\prime 2}}{4}  \, , \nonumber \\
	a_{1}= -a_{2}=a_{6}^{(N)}=2a_{7}=\frac{1}{2} \, , \nonumber \\
	a_{8}^{(N)}= {g_{N}\over2} \, , \nonumber \\
	a_{3}^{ab}= f^{daz} f^{dbz} \, , \nonumber \\
	a_{9}=	\frac{g^{\prime}}{2}   \, .
	\label{coefsigmac}
\end{eqnarray}

Here we must highlight a fundamental point that brings mathematical consistency to this result. The $s_{q}$ operator is easily seen as a zero eigenvalue filtration on the ghost $c^{a}$ of the full $s_{q}+\delta$ nilpotent operator, equivalent to the filtration shown in (\ref{filter}), with $s_{q}$ playing the role of  $s_{0}$ and  $\delta$ playing the role of  $s_{1}$ in that equation. Then, by the fundamental theorem of cohomology already mentioned \cite{Sor95}, the cohomology of $s_{q}+\delta$ is isomorphic to a subspace of the cohomology of $s_{q}$. This means that the independent elements of (\ref{sigmac}) after imposing $\delta$ as a symmetry are in fact independent cocycles of the cohomology of the full BRST operator $s_{q}+\delta$. Their coefficients are then interpreted as independent observables of the theory. Finally, from (\ref{coefsigmac}) we are allowed to conclude that the couplings $g_{N}$ and $g^{\prime}$ are independent observables of the broken phase.

In the list  (\ref{coefsigmac}), we omitted the coefficients $a_{4}$, $a_{5}$ and $D^{abcd}$ . As we mentioned in the beginning of this section, the invariance under $\delta$ imposes that wherever $A^{a}_{\mu}$ appears non-derivated it must be completed in the sense of the Russian-like formula (\ref{VA}). Then, if we limit the counterterm cohomology to the maximum UV dimension of 4, the contributions associated to $a_{4}$, $a_{5}$ and $D^{abcd}$ are excluded as these elements demand extra terms extrapolating this boundary in order to be $\delta$ invariant. In this way, they cannot contribute to the counterterm action. But we guess that once we are in a lower energy regime, possibly beyond a confining scale, these elements can origin contributions to condensates, together with all the independent elements of the cohomology of $s_{q}$ present in (\ref{sigmac}).

Now we can address the question for the gauge fixing after the symmetry breaking. The BRST definition (\ref{sgf}) for the 't Hooft gauge is implemented by using the vacuum operator for this phase

\begin{eqnarray}
S_{gf}&=& (s_{q}+\delta)\int d^{4}x \left(2q^{A}G^{A}+\alpha_Q q^{a}b^{a} +\alpha_N q^i_Nb^i_N+\alpha^\prime q^zb^z\right)\, .
\label{sgfs}
\end{eqnarray}
Here the $G^{A}$, previously given in (\ref{ga}), are rewritten as
\begin{eqnarray}
	G^{i(N)}&=&\partial_{\mu}A^{i(N)}_\mu\nonumber\\
	G^{z}&=&\partial_{\mu}A^{z}_\mu\nonumber\\ 
	G^{a}&=&\partial_{\mu}A^{a}_\mu+ \alpha_Q \mu f^{azb}\chi^{b} \,.
	\label{ga}
\end{eqnarray}
With independent gauge parameters for each sub-group, $\alpha_N$, $\alpha_z$ and $\alpha_Q$, the gauge fixing term turns out to be

\begin{eqnarray}
S_{gf}&=&\int d^{4}x \left( 2b^{i}\partial_{\mu}A^{i\mu} +
2q^{i}_N\left(\partial^2c^i+g_Nf^{ijk}\partial^{\mu}(A^j_{\mu}c^k) + 
\frac{1}{g_N}f^{iab}\partial^{\mu}(A^a_{\mu}c^b)\right)+
2b^{z}\partial_{\mu}A^{z\mu} +\right.\nonumber\\
&&\left. 2q^{z}\left(\partial^2c^z+ \frac{1}{g^\prime}f^{zab}\partial^{\mu}(A^a_{\mu}c^b)\right)+2b^{a}\left(\partial_{\mu}A^{a\mu} +\alpha_{Q}\mu f^{azb}\chi^{b}\right)+2q^{a}\left[\partial^2c^a+ g_Nf^{iab}\partial^{\mu}\left(A^b_{\mu}c^i\right)+\right.\right.\nonumber\\
&&\left.\left.{g^\prime}f^{zab}\partial^{\mu}(A^b_{\mu}c^z)+{g_N}f^{aib}\partial^{\mu}\left(A^i_{\mu}c^b\right)+{g^\prime}f^{azb}\partial^{\mu}(A^z_{\mu}c^b)+
\alpha_{Q}\mu g_Nf^{azb}f^{bcj}(\chi^{c}c^j)\right.\right.\nonumber\\
&&\left.\left.+\alpha_{Q}\mu g^\prime f^{azb}f^{bcz}(\chi^{c}c^z) +\alpha_{Q}\mu g_Nf^{azb}f^{bjc}(\chi^{j}c^c)+\alpha_{Q}\mu g^\prime f^{azb}f^{bzc}(\chi^{z}c^c)
-\alpha_{Q}\mu^2 f^{azb}f^{bcz}c^c\right]\right.+\nonumber\\&&\left.
\alpha_{Q}b^ab^a+ \alpha^\prime b^zb^z+\alpha_{N}b^ib^i\right)\,.
\end{eqnarray}

As promised, closing this section, we will present an explicit calculation on the symmetric symbols $D^{abcd}$ appearing in (\ref{sigmac}). Let us take a toy model with a GUT based on a $SU(3)$ group. The scalar fields are in the adjoint representation, but for convenience we expand them on the Gell-Mann basis of generators of the $su(3)$ algebra  in a fundamental representation \cite{Gel62} (this kind of expansion, in fact, is used in one of the Higgses of the minimal $SU(5)$ GUT for example). We can break the symmetry along the $z=8$ direction, leaving a $SU(2)\times U(1)$ symmetry of the vacuum. In our notation, the indices $i$ will be associated to the first three generators expanding the $su(2)$ subalgebra, and the $a$ will represent the remaining four broken directions. In this particular model, one can show that a possible symmetric contribution may be expanded in the tensors of the algebra
\begin{equation}
 d^{iab}A^{a}_{\mu}A^{b}_{\nu}d^{icd}A^{c\mu}A^{d\nu}=\frac{1}{4}((A^{a}_{\mu}A^{a\mu})^{2}-\frac{4}{3}(f^{8ab}A^{a}_{\mu}A^{b}_{\nu}f^{8cd}A^{c\mu}A^{d\nu})) \, .
 \label{dsu3}
\end{equation}
This relation can be demonstrated from an identity specific for the $su(3)$ case \cite{Que18}, which is possible by the cancellation of the totally symmetric $d^{abcd}$ for $su(N<4)$. Our argument is just to stress that the elements $D^{abcd}$ need to be studied in each case, and in this present simple example they can be discarded as, from (\ref{dsu3}), they can be written in terms of elements already present at (\ref{sigmac}).

\section{Conclusions}

The action (\ref{sigmac}) with (\ref{coefsigmac}) displays the different couplings associated to each group that remains an invariance of the vacuum after the phase transition. As we have shown, they are associated to independent nontrivial cocycles of the broken phase BRST operator $s_{q}+\delta$. Then, following the standard BRST quantization procedure,
these couplings now develop in an independent way with their own renormalization equations in this phase. It is important to recall that
they are tied by the boundary condition (\ref{gs}), i.e., that they all  must converge to the value of the original coupling $g$ at the GUT phase transition scale $M_{G}$. 

Another feature of the solution (\ref{sigmac}) with (\ref{coefsigmac}) for the cohomology problem is that all couplings can be reabsorbed under independent redefinitions of the gauge fields. Physically this associates the independence of each coupling to the acquired independent renormalization of these fields belonging to the different symmetry groups labelled by N that remain as symmetries of the vacuum after the phase transition.

Finally, the presence of the contributions with coefficients $a_{4}$, $a_{5}$ and  $D^{abcd}$ in (\ref{sigmac}) must be emphasized as these objects are not originated from any element already present at the action before the symmetry breaking. They may be interpreted as contributions coming from elements of superior UV dimension of the symmetric phase. They may become relevant in a confined low energy regime, together with all the independent elements present in (\ref{sigmac}). This seems to be a novelty associated to a phase transition process, that can be helpful in the definition of new observables of a broken phase, as condensates for example.

\end{document}